\documentclass[aps,prl,amsmath, twocolumn, groupedaddress, showpacs]{revtex4}
\usepackage{amstext}
\usepackage{graphicx}
\usepackage{amssymb}



\begin{document}

\preprint{}

\title{Catching Time-Reversed Microwave Photons with 99.4\% Absorption Efficiency}

\author{ J. Wenner$^1$}
\author{ Yi Yin$^{1,2}$}
\author{ Yu Chen$^1$}
\author{ R. Barends$^1$}
\author{ B. Chiaro$^1$}
\author{ E. Jeffrey$^1$}
\author{ J. Kelly$^1$}
\author{ A. Megrant$^{1,3}$}
\author{ J. Y. Mutus$^1$}
\author{ C. Neill$^1$}
\author{ P. J. J. O'Malley$^1$}
\author{ P. Roushan$^1$}
\author{ D. Sank$^1$}
\author{ A. Vainsencher$^1$}
\author{ T. C. White$^1$}
\author{ Alexander N. Korotkov$^4$}
\author{ A. N. Cleland$^1$}
\author{ John M. Martinis$^1$}
\email{martinis@physics.ucsb.edu}

\affiliation{$^1$Department of Physics, University of California, Santa Barbara, California 93106, USA}
\affiliation{$^2$Department of Physics, Zhejiang University, Hangzhou 310027, China}
\affiliation{$^3$Department of Materials, University of California, Santa Barbara, California 93106, USA}
\affiliation{$^4$Department of Electrical Engineering, University of California, Riverside, California 92521, USA}

\begin{abstract}
We demonstrate a high efficiency deterministic quantum receiver to convert flying qubits to logic qubits. We employ a superconducting resonator, which is driven with a shaped pulse through an adjustable coupler. For the ideal ``time reversed'' shape, we measure absorption and receiver fidelities at the single microwave photon level of, respectively, 99.41\% and 97.4\%. These fidelities are comparable with gates and measurement and exceed the deterministic quantum communication and computation fault tolerant thresholds.
\end{abstract}


\maketitle

Systems coupling qubits and cavities provide a natural interface between fixed logic qubits and flying photons. They have enabled a wide variety of important advances in circuit quantum optics, such as generating novel photon states \cite{Kirchmair2013, Wang2010, vanLoo2013} and developing a toolbox of quantum devices \cite{Hoi2013,Miranowicz2013, Peropadre2011, Reiserer2013}. Hybrid quantum devices and computers \cite{Wallquist2009, Xiang2013} can be implemented between superconducting coplanar waveguides and micromechanical oscillators or superconducting, spin, and atomic qubits \cite{Schuster2010, Palomaki2013, Majer2007, Hogan2012, Kubo2010}. For such implementations, it can be advantageous to employ deterministic quantum state transfer \cite{Ritter2012, Steffen2013}, which needs highly efficient quantum transmitters and receivers.

\begin{figure}
\includegraphics{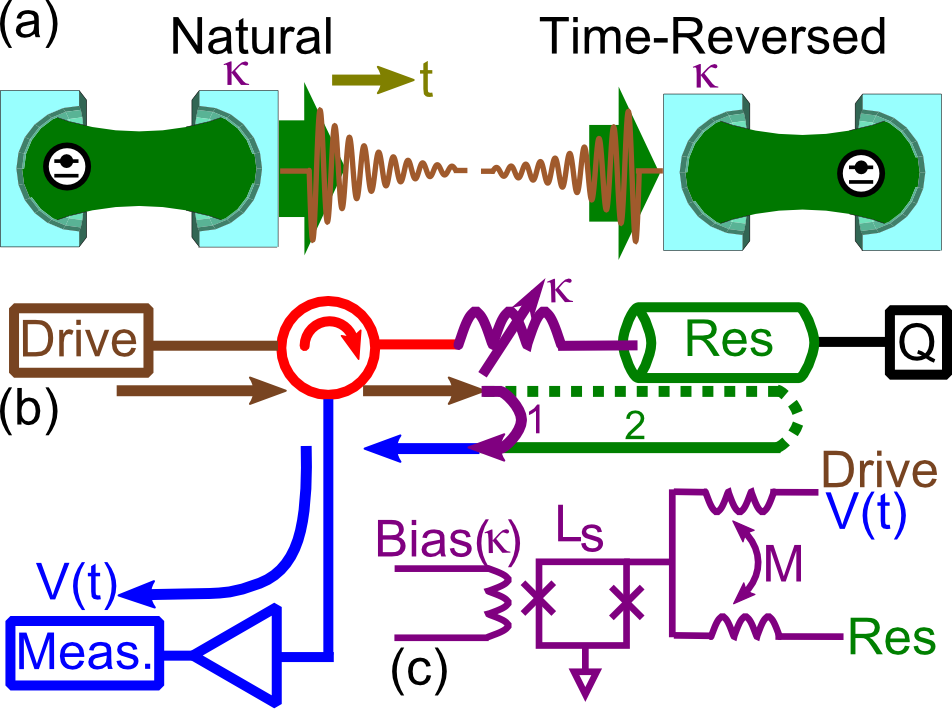}
\caption{\label{FigSchematic} (Color online) Experimental design. (a) Natural decay of a photon state from a cavity, showing exponential decay with energy leakage rate $\kappa$. A time-reversed photon wavepacket is absorbed by a cavity with 100\% efficiency. (b) Experimental setup showing shaped microwave pulse (drive) sent to the cavity (resonator) through a coupler ($\kappa$). An on/off tunable coupler allows separation of the capture, storage and release processes, as well as calibration. An off-chip circulator separates the drive (brown arrows) from the output $V(t)$ (blue arrows), measured with an amplifier and comprised of reflected (purple arrow, 1) and retransmitted (solid green, 2) signals that interfere destructively. The resonator is capacitively coupled to a superconducting phase qubit (Q) for calibration. (c) Schematic of tunable coupler. Coupler consists of two interwoven inductors with negative mutual inductance $M$ and a SQUID with inductance $L_s$ tuned by coupler bias line to vary $\kappa$.}
\end{figure}

For deterministic quantum networks \cite{Kimble2008}, it is particularly challenging to convert flying qubits to stationary qubits, since absorbing naturally shaped emission has a maximum fidelity of only 54\% \cite{Stobinska2009, Wang2011}. Theoretical protocols reaching 100\% efficiency rely upon sculpting the time dependence of photon wavepackets and receiver coupling \cite{Cirac1997, Jahne2007, Korotkov2011}. To accomplish this, recent experiments have developed transmitters with adjustable coupling \cite{Yin2013, Srinivasan2013, Pechal2013}. However, experimental reception fidelities have reached a maximum of only 88\% for optical photons \cite{Bader2013} and 81\% for microwave photons \cite{Palomaki2013}. These are well below fidelities required for fault-tolerant deterministic quantum communication (96\%, \cite{Fowler2010}) and computation (99.4\%, \cite{Fowler2012}).

We demonstrate here a quantum receiver that absorbs and stores photons with a surprisingly high fidelity above these thresholds. We classically drive a superconducting coplanar waveguide resonator through an adjustable coupler, which we use as an on/off switch, with a particularly simple ``time reversed'' photon shape. At the single microwave photon level, we measure an absorption efficiency of 99.4\% and a receiver efficiency of 97.4\%. These efficiencies are comparable to fidelities of good logic gates and measurements \cite{Chow2012, Johnson2012, Geerlings2013}, enabling new designs for quantum communication and computation systems.

\begin{figure*}
\includegraphics{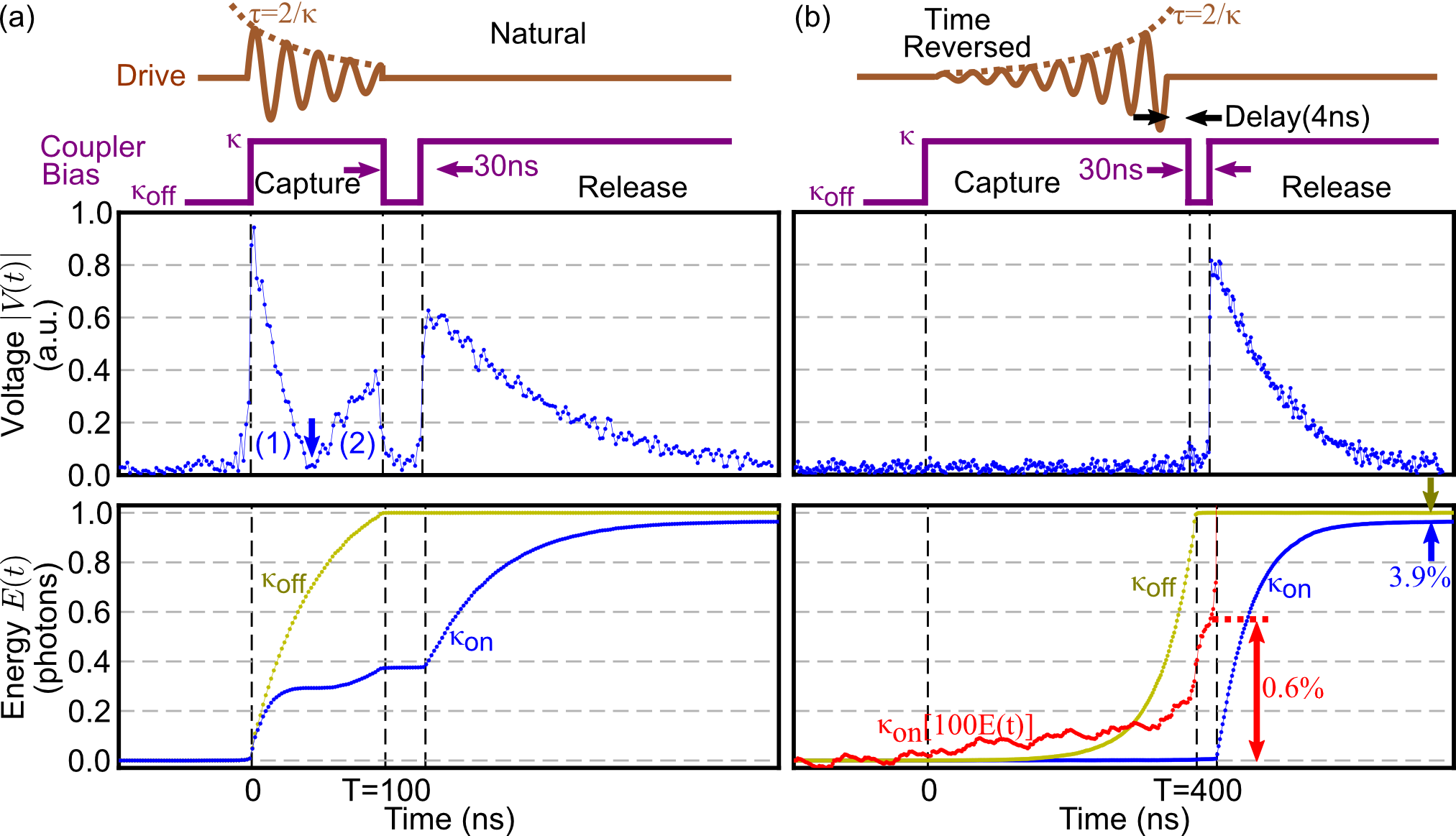}
\caption{\label{FigRawReflection} (Color online) Measurement protocol and data. (Top) Pulse sequence starts by driving the resonator with the coupler on using (a) a natural exponentially decreasing or (b) time-reversed exponentially increasing microwave pulse. Following the drive pulse, we close the coupler for 30\,ns and reopen it to release the resonator energy. The packets have an energy of one photon and an amplitude time constant $\tau=2/\kappa=100$\,ns and are truncated after length $T$=100\,ns (natural) or 400\,ns (time reversed).
(Middle) Measured output voltage versus time. For the natural wavepacket (a), we see an initial reflection (1) and then resonator retransmission (2) in the capture period, followed by the release of the stored energy.  For time-reversed packet (b), little microwave power is observed in the capture period, indicating high efficiency. The complex voltages $V(t)$ are averaged over $3\times10^6$ runs.
(Bottom) Blue ($\kappa_{\mbox{on}}$) curves show measured energy $E(t)$, obtained by subtracting the average noise power from $|V(t)|^2$ and then integrating over time. Gold ($\kappa_{\mbox{off}}$) curves show calibration signal of the reflected incoming packet, with the coupler always off. In (b) the red ($\kappa_{\mbox{on}}[100E(t)]$) curve is a $100\times$ expanded scale, showing small reflected energy. The energy at the end of the capture period, normalized to the total energy at long times, gives the absorption error; we find absorption efficiencies of (61.0$\pm$0.3)\% for the natural and (99.41$\pm$0.06)\% for the time reversed case. Normalizing to the total incident energy (gold, $\kappa_{\mbox{off}}$), we measure a [95.5$\pm$1.2]\% ([97.4$\pm$0.6]\%) storage (receiver) efficiency for the time reversed wavepacket.
}
\end{figure*}

The protocol we implement relies on time-reversal symmetry \cite{Jackson, Cirac1997} between resonator energy absorption and emission. A resonator emits a photon wavepacket which naturally decays exponentially in amplitude, as shown in Fig.\,\ref{FigSchematic}(a). By time reversal symmetry, the resonator will thus absorb an exponentially increasing wavepacket with perfect efficiency \cite{Stobinska2009, Wang2011, Heugel2010}.

More physically, the resonator perfectly absorbs a travelling wave packet if destructive interference occurs between signals reflected off the coupler and retransmitted (leaked) out of the resonator. This is readily obtained for an exponentially increasing wavepacket, as calculated both classically \cite{Heugel2010, Korotkov2011, Supp} and quantum mechanically \cite{Wang2011, Stobinska2009}, since the reflected and retransmitted signals increase in time together with opposite phase. For perfect destructive interference during the entire pulse, one must match the frequencies and set the drive amplitude time constant $\tau$ to $2/\kappa$ for coupling leakage rate $\kappa$. Under these conditions, the absorption efficiency equals $1-e^{-\kappa T}$ for a pulse of duration $T$ \cite{Supp}. The absorption efficiencies reach 99.4\% for $\kappa T\geq5.3$ and approach unity as $T\rightarrow\infty$. Using this idea, other protocols also permit perfect cancelation by temporally varying both the wavepacket and $\kappa$ \cite{Korotkov2011}.

Imperfect destructive interference between the reflected and retransmitted waves results in lower absorption efficiencies. The maximal absorption efficiency is only 81.45\% \cite{Supp} for a drive pulse with rectangular or exponentially decreasing amplitude because initially the reflected signal has greater amplitude than the retransmitted signal, while at long times, retransmission dominates. Reflections also do not cancel out retransmission if the resonator couples significantly to modes besides the drive; this results in a decreased efficiency as seen in Refs.\,\cite{Aljunid2013, Zhang2012, Palomaki2013}.

To experimentally achieve the strong coupling necessary for complete destructive interference, we employ a tunable inductive coupler, shown in Fig.\,\ref{FigSchematic}(b,c), through which we drive a 6.55\,GHz superconducting coplanar waveguide resonator. The coupling is given by a mutual inductance modulated by a tunable SQUID (superconducting quantum interference device) inductance \cite{Yin2013} and is calibrated using a superconducting phase qubit \cite{Supp}. The resulting coupling can be tuned from negative couplings through zero (off) to $+1/(20$\,ns) in a few nanoseconds; as the drive pulse is similarly tuned in a few nanoseconds, the coupling can be adjusted with the drive pulse to ensure reflection cancellation. We choose a coupling $\kappa=+1/(50$\,ns) for resonator driving and retransmission, which dominates over the intrinsic resonator loss $\kappa_i\simeq1/(3\,\mu$s). By turning off the coupler, the absorbed energy is stored instead of immediately retransmitted.

\begin{figure}
\includegraphics{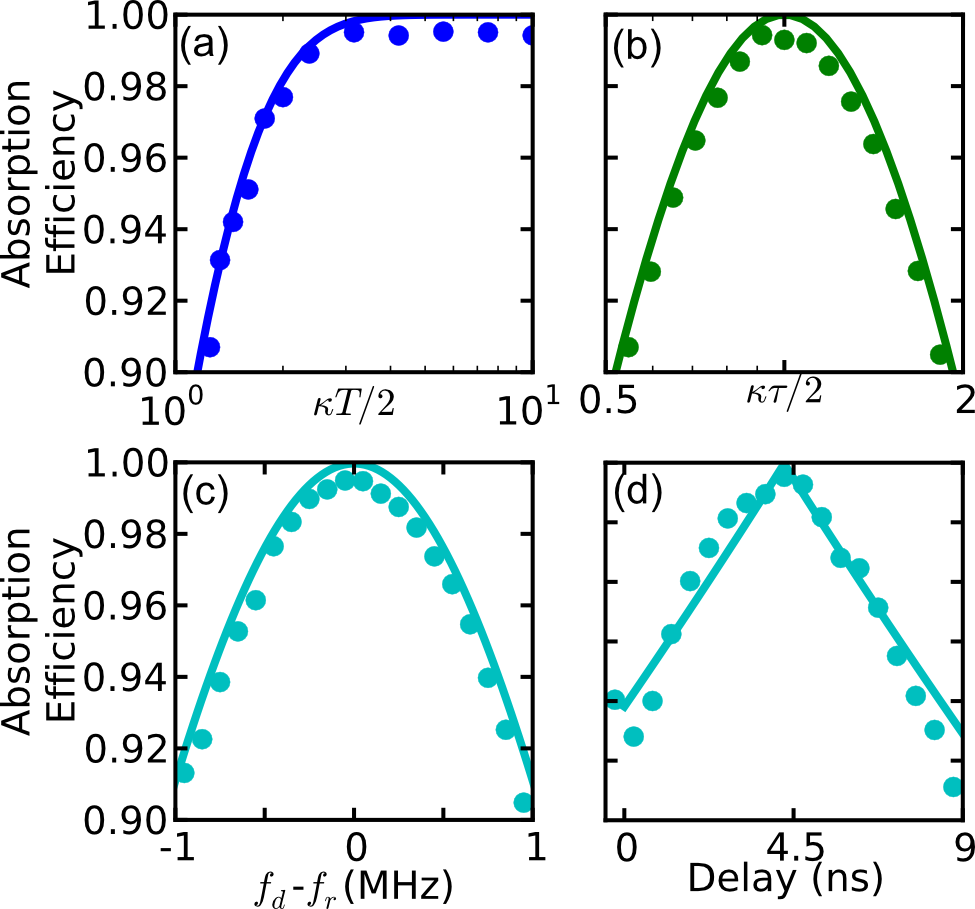}
\caption{\label{FigTuneup} (Color online) Tuning up 99.4\% absorption efficiency. As shown in the pulse sequence of Fig.\,\ref{FigRawReflection}(a), the four parameters which must be tuned up precisely are (a) the pulse length $T$, (b) time constant $\tau$, (c) detuning of the drive frequency $f_d$ relative to the resonator frequency $f_r$, and (d) time delay of closing coupler after stopping resonator drive. The absorption efficiencies are maximized for an exponentially increasing pulse by $\tau=2/\kappa$, $T\geq5.3/\kappa$, $f_d=f_r$, and 4.5\,ns delay. In all cases the experimental absorption efficiencies (points) fall off as expected theoretically (lines \cite{Supp}). Data are for single photon drives with $\kappa=1/(50$\,ns) and (a) $\tau=2/\kappa$, (b) $T=20/\kappa$, (c)-(d) $\tau=2/\kappa$ and $T=8/\kappa$, as per the colored slices in Fig.\,\ref{FigGridEfficiencies}(e). Uncertainties are $\leq0.14\%$ \cite{Supp}.}
\end{figure}

We drive the resonator with a shaped pulse and measure the output voltage $V(t)$ versus time $t$ to characterize the destructive interference. We generate a classical single-photon drive pulse using heterodyne mixing with an arbitrary waveform generator \cite{Supp} and concurrently set the coupling to $\kappa=+1/(50$\,ns) to capture the drive energy. Upon stopping the drive, we idle the coupler at $\kappa_{\mbox{off}}=0$ for 30\,ns and then reset the coupling to $\kappa$ to release the energy [see pulse sequence in Fig.\,\ref{FigRawReflection}]. During the entire sequence, we measure the complex $V(t)$ using two-channel heterodyne detection near the resonator frequency \cite{Supp}, averaging over $3\times10^6$ repetitions.

When the reflection and retransmission interfere destructively, $V(t)$ is comparable to the noise prior to the drive. This occurs only around 50\,ns for an exponentially decreasing drive pulse [arrow in Fig.\,\ref{FigRawReflection}(a)]; elsewhere, either reflection [(1) in the middle panel] or retransmission (2) dominate. In contrast, the destructive interference lasts the entire drive for a properly shaped exponentially increasing pulse [Fig.\,\ref{FigRawReflection}(b)], implying high absorption efficiency.

We quantify this interference by the energy absorption efficiency. The energy measured through time $t$ is proportional to $E(t)=\int_0^t [|V(t')|^2-N]\,dt'$ for average noise power $N$ [see bottom panel, Fig.\,\ref{FigRawReflection}]. The absorbed energy $E_{\mbox{abs}}$ is the difference between the total drive energy $E(\infty)$ and the near-constant energy $E(\mbox{idle})$ at the idle. The absorption efficiency $E_{\mbox{abs}}/E(\infty)$ equals (99.41$\pm$0.06)\% for the exponentially increasing drive but (61.0$\pm$0.3)\% for the natural exponentially decreasing drive. This high absorption efficiency is independent of both the number of repetitions over which $V(t)$ is averaged and $E(\infty)$ for $E_{\mbox{abs}}=(0.5\rightarrow50)$ photons \cite{Supp}. Note that we calibrate the energy scale by using the qubit to measure $E_{\mbox{abs}}$ \cite{Supp}.

To achieve this high 99.4\% absorption efficiency, we tune the pulse length $T$, exponential time constant $\tau$, drive frequency $f_d$, and the timing of closing the coupler (see Fig.\,\ref{FigTuneup}). For $\tau=2/\kappa$, longer pulse lengths correspond to higher absorption efficiencies, reaching 99.4\% for $T\geq6/\kappa$ [Fig.\,\ref{FigTuneup}(a)]. If $\tau$ is varied at $T=20/\kappa$ [Fig.\,\ref{FigTuneup}(b)], the absorption efficiency reaches the maximal 99.4\% within 10\% of $\tau=(2/\kappa)$ but falls to 90\% for $\tau/(2/\kappa)=2,1/2$. In both cases, the efficiency falls off as expected.

We also tune the drive frequency $f_d$ about the resonator frequency $f_r$ for a $T=8/\kappa$, $\tau=2/\kappa$ exponentially increasing drive pulse [Fig.\,\ref{FigTuneup}(c)]. The absorption efficiency is maximized for $f_d=f_r$ and is at least 90\% within $\pm$1\,MHz of $f_r$. According to theory, achieving appreciable absorption efficiency requires $f_d=f_r$ to within a linewidth $\kappa/2\pi=$3\,MHz \cite{Supp}.

To achieve the 99.4\% absorption efficiency, we must delay closing the coupler relative to turning off the drive even after calibrating the coupler-resonator timing \cite{Supp}. The absorption efficiency is reduced by 10\% when the delay differs by 3.5\,ns from the optimal 4.5\,ns [Fig.\,\ref{FigTuneup}(d)]. The efficiency decreases since the entire drive is reflected if the coupler is closed too early and some of the captured energy is retransmitted if the coupler is closed too late. The scaling is linear for delays longer (shorter) than the optimum due to the sharpness of turning off the drive (closing the coupler). We observe experimental deviations from the linear slope for shorter delays, so our coupler pulse shaping is nonideal, possibly explaining why the offset is required.

\begin{figure}
\includegraphics{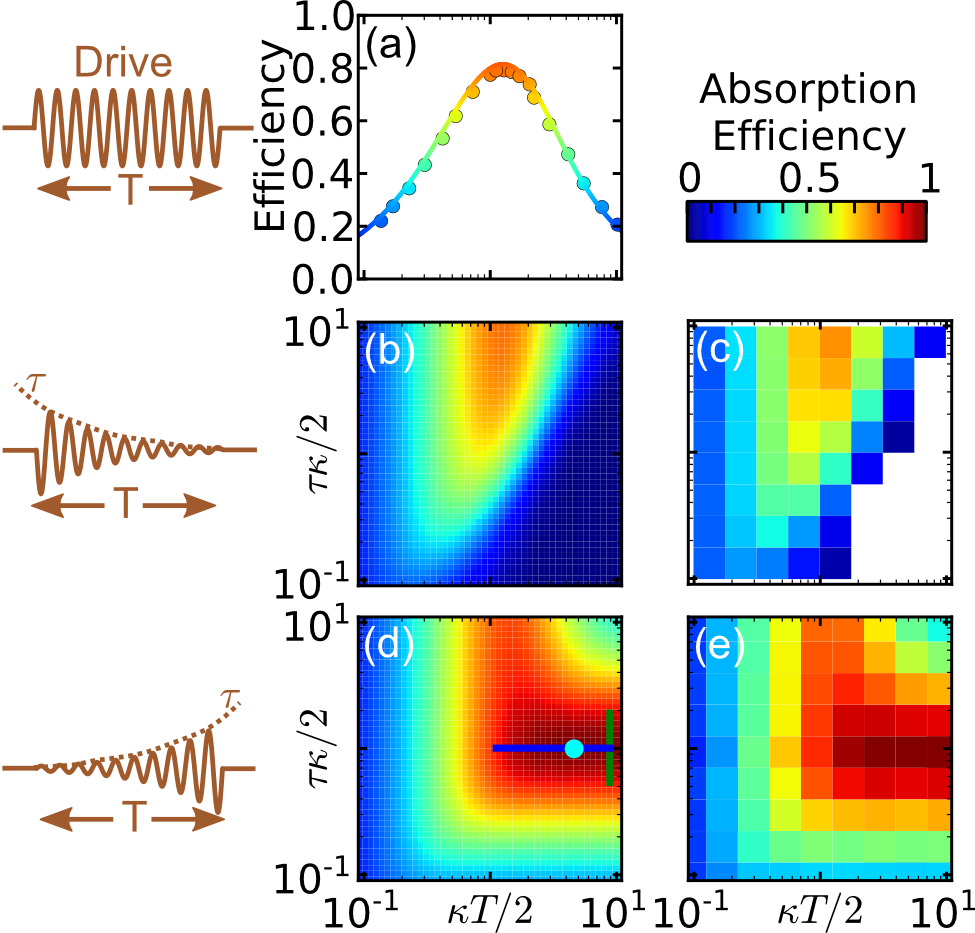}
\caption{\label{FigGridEfficiencies} (Color online) Absorption efficiencies vs pulse shape. Theoretical and experimental absorption efficiencies (color, grayscale) are plotted versus pulse length $T$ and exponential time constant $\tau$. Experimental data are for single photon drives with couplings (a,c) $\kappa=1/(40\,$ns) and (e) $\kappa=1/(50\,$ns) and have uncertainties $\leq0.4\%$ \cite{Supp}. Theory has no fit parameters. (a) Rectangular drive pulse (theory and experiment). Absorption effiencies are maximized at 79\% (81.5\% theory) for $T\approx2.5/\kappa$. (b,c) For truncated exponentially decreasing drive pulse, theoretical (b) and experimental (c) absorption efficiencies are maximized as $\tau\rightarrow\infty$, the rectangular pulse limit, with efficiencies similar to (a). (d,e) For truncated exponentially increasing drive pulse, theoretical (d) and experimental (e) absorption efficiencies are maximized for $\tau=2/\kappa$ and $T\geq6/\kappa$ at 99.4\% (100\% theory).}
\end{figure}

To demonstrate the necessity of appropriate pulse shaping, we measured the absorption efficiencies for rectangular [Fig.\,\ref{FigGridEfficiencies}(a)], truncated exponentially decreasing [Fig.\,\ref{FigGridEfficiencies}(c)], and exponentially increasing drive pulses [Fig.\,\ref{FigGridEfficiencies}(e)] versus pulse length $T$ and exponential time constant $\tau$. All three pulse shapes have similar shapes and hence absorption efficiencies for $\tau>10\,T$. For rectangular pulses, the maximal absorption efficiency is $\sim79\%$ compared to the predicted 81.5\%; this limit also provides the maximal absorption efficiencies for exponentially decreasing pulses. The efficiency is reduced because no initial excitation exists for which retransmission can cancel the initial reflections [see (1) in middle panel, Fig.\,\ref{FigRawReflection}(a)]. In addition, these experimental efficiencies agree with the theoretical efficiencies [Fig.\,\ref{FigGridEfficiencies}(a),(b),(d)], to within $\sim4$\%, with possible error sources including calibration and pulse shaping errors. Both theory and experiment show that the only pulses with 99.4\% absorption efficiencies are exponentially increasing pulses with $\tau=2/\kappa$ and $T\geq6/\kappa$.

However, the absorption efficiencies neglect intrinsic resonator losses $\kappa_i$. To measure this effect, we drive the resonator with the coupler off ($\kappa_{\mbox{off}}$) and measure the total reflected energy $E_{\mbox{off}}$. We compare this to the total measured energy $E_{\mbox{on}}$ when we drive the resonator at $\kappa$. The fraction of energy not lost is $E_{\mbox{on}}/E_{\mbox{off}}$=96.1\% [Fig.\,\ref{FigRawReflection}(b), bottom panel]. The storage efficiency for the entire process, (95.5$\pm$1.2)\%, is $E_{\mbox{on}}/E_{\mbox{off}}$ times the absorption efficiency. However, this efficiency includes losses from $\kappa_i$ during both the capture and release phases. During just the drive, we expect to keep $\kappa/(\kappa+\kappa_i)=98.4\%$ of the energy \cite{Supp} and actually keep approximately $\sqrt{E_{\mbox{on}}/E_{\mbox{off}}}=98.1\%$ assuming near-perfect absorption \cite{Palomaki2013}. This fraction times the absorption efficiency is the receiver efficiency, which equals (97.4$\pm$0.6)\% for the optimal pulse.

In conclusion, we have demonstrated 99.4\% photonic energy absorption when driving a superconducting coplanar waveguide resonator with exponentially-increasing pulses. With this time-reversed drive, the reflected and retransmitted signals are effectively cancelled. The resonator receives 97.4\% of the drive energy since the resonator-drive coupling dominates over intrinsic loss mechanisms. As resonators have been developed with coherence times $1/\kappa_i=45\,\mu$s \cite{Megrant2012}, we should be able to increase this receiver efficiency to above 99\%. Hence, deterministic quantum state transfer between macroscopically separated cavities is possible with high fidelity.

\section{Acknowledgments}
Devices were made at the UC Santa Barbara Nanofabrication Facility, part of the NSF-funded National Nanotechnology Infrastructure Network. This research was funded by the Office of the Director of National Intelligence (ODNI), Intelligence Advanced Research Projects Activity (IARPA), through Army Research Office grant W911NF-10-1-0334. All statements of fact, opinion or conclusions contained herein are those of the authors and should not be construed as representing the official views or policies of IARPA, the ODNI, or the U.S. Government.

\end{document}


\title{Supplement to ``Catching Shaped Microwave Photons with 99.4\% Absorption Efficiency''}

\author{ J. Wenner$^1$}
\author{ Yi Yin$^{1,2}$}
\author{ Yu Chen$^1$}
\author{ R. Barends$^1$}
\author{ B. Chiaro$^1$}
\author{ E. Jeffrey$^1$}
\author{ J. Kelly$^1$}
\author{ A. Megrant$^{1,3}$}
\author{ J. Mutus$^1$}
\author{ C. Neill$^1$}
\author{ P. J. J. O'Malley$^1$}
\author{ P. Roushan$^1$}
\author{ D. Sank$^1$}
\author{ A. Vainsencher$^1$}
\author{ T. C. White$^1$}
\author{ Alexander N. Korotkov$^4$}
\author{ A. N. Cleland$^1$}
\author{ John M. Martinis$^1$}

\affiliation{$^1$Department of Physics, University of California, Santa Barbara, California 93106, USA}
\affiliation{$^2$Department of Physics, Zhejiang University, Hangzhou 310027, China}
\affiliation{$^3$Department of Materials, University of California, Santa Barbara, California 93106, USA}
\affiliation{$^4$Department of Electrical Engineering, University of California, Riverside, California 92521, USA}

\renewcommand{\thefigure}{S\arabic{figure}} 
\renewcommand{\thetable}{S\arabic{table}} 
\renewcommand\thepage {S\arabic{page}}
\renewcommand\theequation {S\arabic{equation}}
\setcounter{figure}{0}
\setcounter{table}{0}
\setcounter{page}{1}
\setcounter{equation}{0}

\maketitle
\section{Theoretical Capture Efficiencies}

Here, we calculate the reflected signal and receiver efficiency for tunably-coupled resonator driven with several different drive pulse shapes. For a time constants near $2/\kappa$, exponentially increasing drive pulses are more efficient than rectangular and exponentially decreasing pulses. We then show that optimal efficiencies require that the resonator have zero intrinsic loss and be driven at the resonance frequency.

\subsection{Transmission Coefficients}

\begin{figure}
\includegraphics{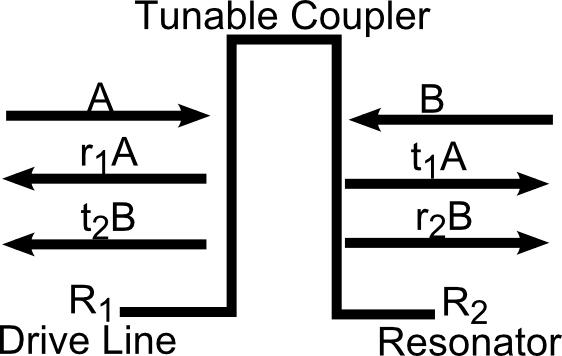}
\caption{\label{FigNotation}
Transmission and reflection coefficient notation. An $R_1\simeq50\,\Omega$ drive transmission line is coupled to an $R_2\simeq80\,\Omega$ resonator via a tunable coupler (barrier). $A$ is the incoming signal from the drive line, whereas $B$ is the signal reaching the coupler from the resonator. The coupler has reflection coefficient $\mathbf{r_1}$ on the drive side, reflection coefficient $\mathbf{r_2}$ on the resonator side, transmission coefficient from the drive line $\mathbf{t_1}$, and transmission coefficient from the resonator $\mathbf{t_2}$.}
\end{figure}

Consider a resonator driven through a tunable coupler with transmission and reflection coefficients defined in Fig.\,\ref{FigNotation}. These coefficients are related by \cite{Korotkov2011}
\begin{equation}
\mathbf{t_1} = \frac{R_2}{R_1}\mathbf{t_2},
|\mathbf{t_1}|^2\frac{R_1}{R_2}+|\mathbf{r}|^2 = 1,
\mathbf{r_2}=-\mathbf{r_1^*}\frac{\mathbf{t_1}}{\mathbf{t_1^*}},
\label{EqT1T2RAll}
\end{equation}
where $R_1\simeq50\,\Omega$ ($R_2$) is the drive (resonator) impedance, $\mathbf{t_1}$ ($\mathbf{t_2}$) is the transmission coefficient into (from) the resonator, and $\mathbf{r_1}$ ($\mathbf{r_2}$) is the reflection coefficient on the drive (resonator) side, where $|\mathbf{r_1}|=|\mathbf{r_2}|=|\mathbf{r}|$.

The transmission coefficients are related to the coupling quality factor $Q_c$ by
\[
Q_c=\frac{\omega\tau_{rt}}{|\mathbf{t_2}|^2}\frac{R_1}{R_2},
\]
where $\tau_{rt}$ is the ratio of the resonator energy to the travelling wave power and $\omega/2\pi=6.55$\,GHz is the resonator frequency. For a $\lambda/4$ coplanar waveguide resonator \cite{Korotkov2011},
\begin{equation}
\tau_{rt}\approx\pi/\omega.
\label{EqTrt}
\end{equation}
With our device, $\kappa\tau_{rt}\ll1$ as the coupler energy decay rate $\kappa$ is within the range $[1/(3\,\mu\mbox{s}),1/(50\,\mbox{ns})]$. According to \cite{Wang2009}, $\kappa$ is given by $Q_c=\omega/\kappa$, so
\begin{equation}
\frac{|\mathbf{t_1}|^2}{\tau_{rt}}=\kappa\frac{R_2}{R_1}.
\label{EqKappa}
\end{equation}
Note that $|\mathbf{t_1}|^2\ll1$ as the coupler is near the grounded end of the resonator.

\subsection{Absorption Efficiencies}

Suppose that two signals reach the coupler at time $t$: an incoming drive $A(t)$ and a signal $B(t)$ from the resonator (see Fig.\,\ref{FigNotation}). Then, $B(t)$ is described by \cite{Korotkov2011}
\begin{equation}
\dot{B} = \frac{\mathbf{r_2}e^{i\phi}-1}{\tau_{rt}}B + \frac{\mathbf{t_1}e^{i\phi}}{\tau_{rt}}A - \frac{1}{2T_1}B,
\label{EqBDot}
\end{equation}
where $T_1$ is the intrinsic resonator energy decay time and $\phi=\tau_{rt}\delta\omega-\mbox{arg}(\mathbf{r_2})$ arises from a detuning $\delta\omega$ between the drive and resonator frequencies. Note that 
\[
e^{i\phi} = \frac{\mathbf{r_2^*}}{|\mathbf{r}|} e^{i\tau_{rt}\delta\omega},
\]
so Eq.\,(\ref{EqBDot}) simplifies to
\begin{equation}
\dot{B} = \left(-\frac{\kappa}{2}B+\frac{\mathbf{t_1}}{\tau_{rt}}\frac{\mathbf{r_2^*}}{|\mathbf{r}|}A\right) - \frac{1}{2T_1}B + \delta\omega\left(iB+\mathbf{t_1}\frac{\mathbf{r_2^*}}{|\mathbf{r}|}A\right)
\label{EqFullBDot}
\end{equation}
with Eq.\,(\ref{EqKappa}) and $\kappa\tau_{rt},\tau_{rt}\delta\omega,|\mathbf{t_1}|^2\ll1$. Here, the time-dependence is contained in $A$, $B$, and potentially $x$.

We initially consider a simple case where the coupling is time-independent, the drive is on resonance, and the resonator intrinsic quality factor is infinite. In this case, solving for $B(t)$ gives
\begin{equation}
B(t) = B(0)e^{-\kappa t/2}+e^{-\kappa t/2}\frac{\mathbf{t_1}}{\tau_{rt}}\frac{\mathbf{r_2^*}}{|\mathbf{r}|}\int_0^{t}A(t')e^{\kappa t'/2}dt'.
\label{EqB}
\end{equation}

The power in the resonator is $|B(t)|^2/2R_2$. The energy $E_{res}$ in the resonator is thus $|B(t)|^2\tau_{rt}/2R_2$ from the definition of $\tau_{rt}$. If the resonator is initially unpopulated,
\begin{equation}
E_{res} = \frac{1}{2R_1}\kappa e^{-\kappa T}\left|\int_0^TA(t)e^{\kappa t/2}dt\right|^2
\label{EqEnergyRes}
\end{equation}
after a time $T$. The total energy $E_{tot}$ equals the integral
\begin{equation}
E_{tot} = \frac{1}{2R_1}\int_0^{T'}|A(t)|^2dt
\label{EqEnergyTot}
\end{equation}
of the drive power $|A(t)|^2/2R_1$ applied for $t\in(0,T')$. The ratio $E_{res}/E_{tot}$ is defined to be the receiver efficiency; note the factors $1/2R_1$ cancel.

\subsection{Rectangular Pulses}

Suppose the resonator is driven by a rectangular pulse, $A(t)=A_0$ for $0\leq t\leq T=T'$. Then, the receiver efficiency is
\begin{equation}
\frac{E_{res}}{E_{tot}} = \frac{4}{\kappa T}\left(1-e^{-\kappa T/2}\right)^2,
\label{EqRectangle}
\end{equation}
which is plotted in Fig.\,4(a) of the main text. This efficiency reaches a maximal value of 81.4529\% when $T=2.51286/\kappa$.

\begin{figure}
\includegraphics{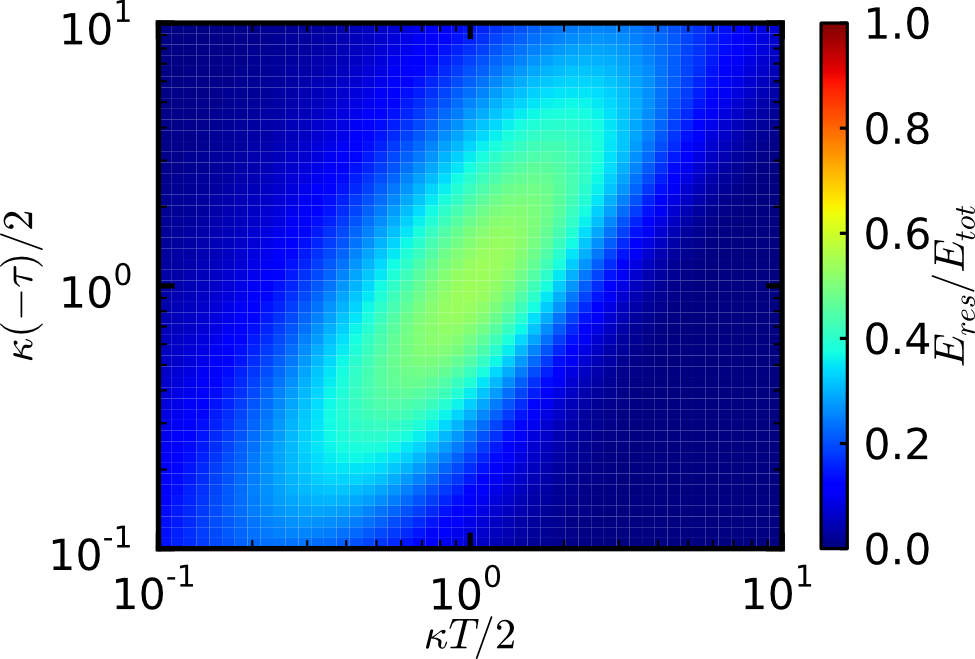}
\caption{\label{FigDecrease}
Receiver efficiency for infinite exponentially decreasing drive. If the drive is infinitely long ($T'\rightarrow\infty$) but the coupler closes at $T$, the receiver efficiency is maximized when $-\tau=T=2/\kappa$ and is then 54.1\%.}
\end{figure}

\subsection{Exponential Pulses}

Suppose that the resonator is driven by an exponential pulse, $A(t)=A_0e^{t/\tau}$ for $0\leq t\leq T'$ and the coupler closed at $t=T$. Then, the receiver efficiency is
\begin{equation}
\frac{E_{res}}{E_{tot}} = \frac{4\left(\frac{2}{\kappa\tau}\right)\left(e^{T/\tau}-e^{-\kappa T/2}\right)^2} {\left(1+\frac{2}{\kappa\tau}\right)^2\left(e^{2T'/\tau}-1\right)}.
\label{EqExpChopped}
\end{equation}
This is true for both exponentially decreasing ($\tau<0$) and exponentially increasing ($\tau>0$) pulses.

The case $\tau<0,T'\rightarrow\infty$ corresponds to a source excitation which naturally decays via static coupling. Here, the receiver efficiency reduces to
\begin{equation}
\frac{E_{res}}{E_{tot}} = \frac{4\left(\frac{2}{\kappa(-\tau)}\right)\left(e^{-T/(-\tau)}-e^{-\kappa T/2}\right)^2} {\left(1-\frac{2}{\kappa(-\tau)}\right)^2}
\label{EqDecreaseUnchopped}
\end{equation}
and is plotted in Fig.\,\ref{FigDecrease}. It is maximized when $-\tau=T=2/\kappa$ with an efficiency of $4/e^2=54.134\%$.

\begin{figure}
\includegraphics{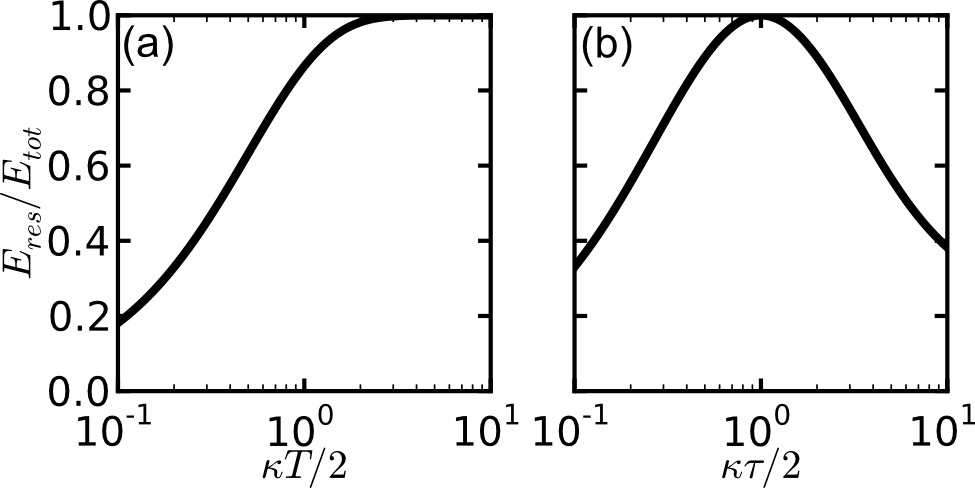}
\caption{\label{FigIncrease}
Receiver efficiency for exponentially increasing drive. The efficiency is shown versus (a) pulse length $T$ ($\tau=2/\kappa$) and (b) time constant $\tau$ ($T\rightarrow\infty$). The efficiency is maximized for $\tau=2/\kappa$, $T\rightarrow\infty$. The regime for efficiencies $>90\%$ is shown in Fig.\,3(a),(b) of the main text. }
\end{figure}

For the $\tau<0,T'=T$ case, a truncated natural drive, the efficiency is plotted in Fig.\,4(b) of the main text. The efficiency is continuous even at $\tau=-2/\kappa$ where
\[
2R_1E_{res} = A_0^2 T^2\kappa e^{-\kappa T}.
\]
For a constant pulse length, the efficiency is maximized in the limit $\tau\rightarrow-\infty$. This limit corresponds to a rectangular pulse, and $E_{res}/E_{tot}$ reduces to the rectangular value. For a constant time constant, the efficiency is maximized at pulse lengths which increase asymptotically as $\tau\rightarrow-\infty$ to the rectangular maximum of $2.51286/\kappa$. In addition, the efficiency always approaches zero for infinite pulse lengths as the excitation has infinite time to decay. 

\begin{figure}
\includegraphics{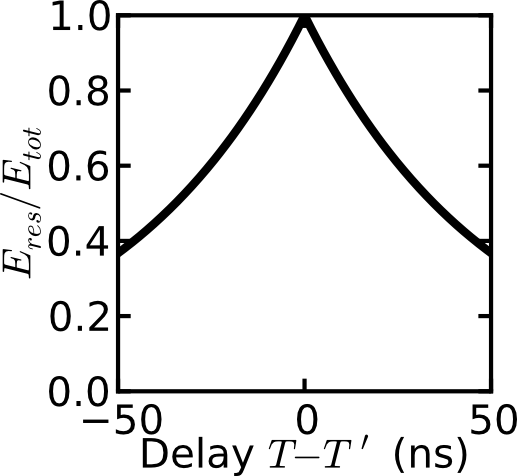}
\caption{\label{FigDelay}
Receiver efficiency versus offset between closing coupler ($T$) and stopping drive ($T'$). The efficiency is shown for an exponentially increasing pulse with $\tau=2/\kappa$, $T=8/\kappa$, $\kappa=1/(50$\,ns). The regime for efficiencies $>90\%$ is shown in Fig.\,3(d) of the main text. The exponential decay for $T<T'$ is due to the particular drive pulse shape, while that for $T'<T$ is universal.}
\end{figure}

\begin{figure*}
\includegraphics{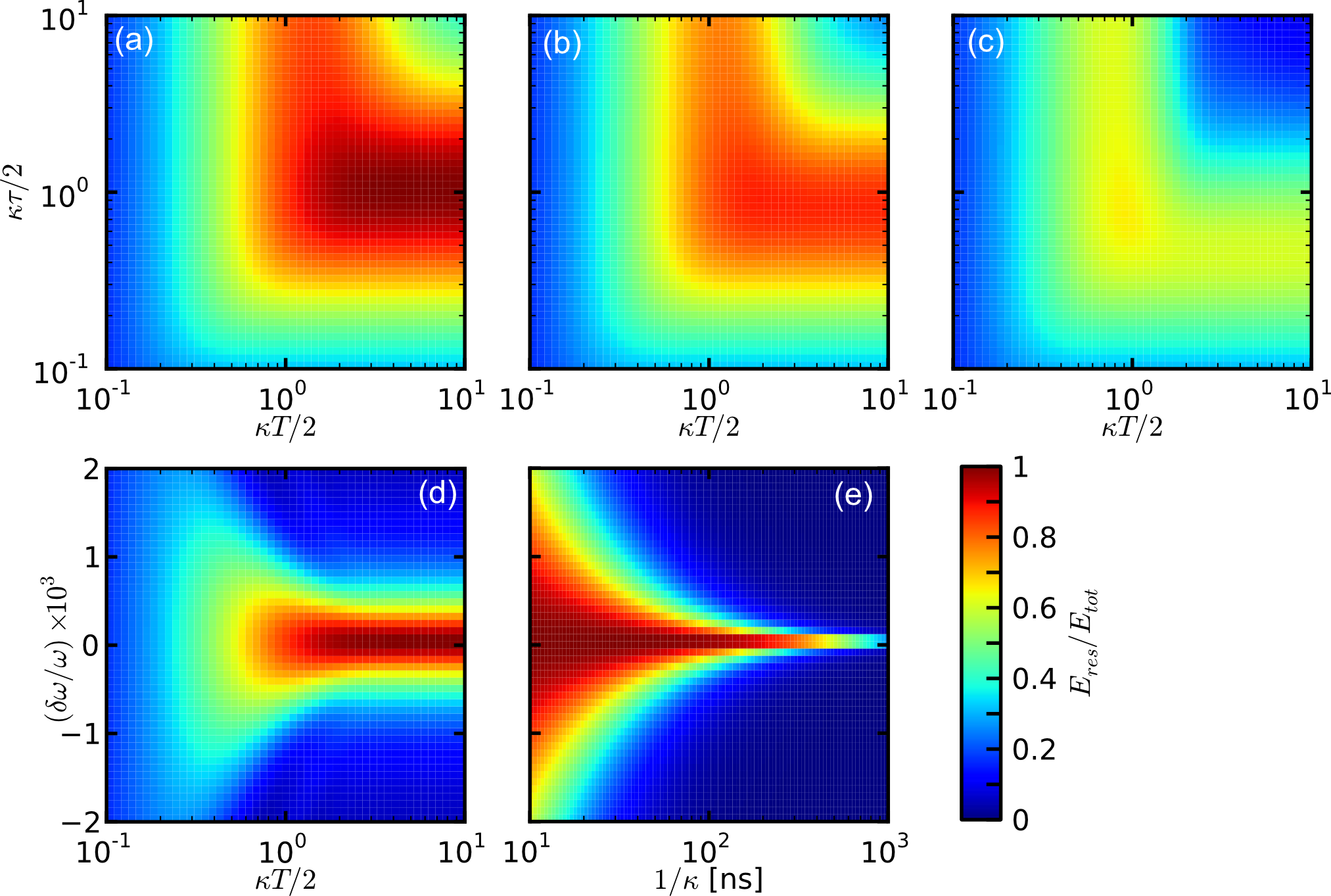}
\caption{\label{FigFreqIncrease}
Effect of detuned drive frequency on receiver efficiency. Unless otherwise stated, theory is for the experimental system with frequency $\omega/2\pi=6.55$\,GHz, coupling $\kappa=1/(50$\,ns), pulse length $T=8/\kappa$, and exponential time constant $\tau=2/\kappa$. (a)-(c) The receiver efficiency is plotted vs $T$ and $\tau$ for three fractional detunings $\delta\omega/\omega$: (a) 0, (b) 0.0002 [1.3\,MHz], and (c) 0.0005 [3.3\,MHz]. (d) The receiver efficiency is plotted vs $T$ and $\delta\omega/\omega$. The efficiency is maximized when the drive and resonator frequencies are equal with the frequency width effectively independent of $T$ for $\kappa T>3$. (e) The receiver efficiency is plotted vs $1/\kappa$ and $\delta\omega/\omega$ for fixed $\kappa T$, $\kappa\tau$. The frequency width is proportional to $\kappa$.}
\end{figure*}

An exponentially increasing ($\tau>0,T'=T$) drive pulse is ideal for static coupling; the receiver efficiency Eq.\,(\ref{EqExpChopped}) is plotted in Fig.\,4(d) of the main text. As predicted \cite{Korotkov2011}, the efficiency for constant pulse lengths is maximized when $\tau=2/\kappa$, where it reduces to $1-e^{-\kappa T}$ and so asymptotically approaches unity as $T\rightarrow\infty$ (see Fig.\,\ref{FigIncrease}(a)). For a fixed time constant, the efficiency is maximized at $T\rightarrow\infty$ for $\tau\leq2/\kappa$ and at lengths asymptotically approaching the rectangular limit as $\tau\rightarrow\infty$. At such pulse lengths, the efficiency is greater than rectangular limit of 81.5\% for $\tau>0.8/\kappa$. For infinite duration pulses, the efficiency simplifies to
\[
\frac{E_{res}}{E_{tot}} = \frac{4\left(\frac{\kappa\tau}{2}\right)} {\left(1+\frac{\kappa\tau}{2}\right)^2},
\]
which is greater than 81.5\% for $0.8<\kappa\tau<5$, as shown in Fig.\,\ref{FigIncrease}(b).

\subsection{Coupler Closing Delay}

Now suppose that the coupler is closed and the drive stopped at slightly different times $T$ and $T'$, respectively. If the drive is stopped first ($T'<T$), the receiver efficiency is decreased by a factor of $e^{-\kappa(T-T')}$ according to Eq.\,(\ref{EqB}), regardless of the pulse shape. If the coupler is closed first ($T<T'$), the receiver efficiency is changed by a factor which depends on the particular drive pulse. For an exponential pulse, the receiver efficiency is reduced by a factor of
\[
\left(\frac{e^{T/\tau}-e^{-\kappa T/2}}{e^{T'/\tau}-e^{-\kappa T'/2}}\right)^2,
\]
as shown in Fig.\,\ref{FigDelay}.

\subsection{Reflections - Destructive Interference}

The basis for these high absorption efficiencies is destructive interference between the reflection $\mathbf{r_1}A$ and retransmission $\mathbf{t_2}B$ signals. This requires opposite phases for these signals. Since the phase of $A$ adds to the phase of $B$, we assume without loss of generality that $A$ is positive. The phase of the reflection signal is then given by $\mathbf{r_1}$, which by Eq.\,(\ref{EqT1T2RAll}) is the phase of $-\mathbf{r_2^*}\mathbf{t_1^2}$. The phase of the retransmission signal is given by the phase of $\mathbf{t_2}$, which by Eq.\,(\ref{EqT1T2RAll}) is the phase of $\mathbf{t_1}$, and the phase of $B$, which by Eq.\,(\ref{EqB}) is the phase of $\mathbf{t_1r_2^*}$. Thus, the reflection and retransmission signals are always opposite in phase.

\subsection{Drive Frequency}

We now start to consider the effects of relaxing our simplifying assumings. First, we detune the drive frequency by a small $\delta\omega$ from the resonator frequency $\omega$ while we still assume that the resonator is lossless and the coupling is static. The capture efficiencies can then be solved with Eq.\,(\ref{EqFullBDot}). The resonator frequency is included here not only in $\delta\omega$ but also $\tau_{rt}$ according to Eq.\,(\ref{EqTrt}).

For the exponentially increasing pulse, a nonzero detuning reduces the maximum receiver efficiency. As the detuning increases, not only is the maximum receiver efficiency reduced but the values of $\tau$ and $T$ which maximize the efficiency are reduced as shown in Fig.\,\ref{FigFreqIncrease}(a)-(c). For the ratio of the efficiency to the efficiency with $\delta\omega=0$, the frequency width of the peak is independent of $T$ for $\tau=2/\kappa$ and $T\geq3/\kappa$ [Fig.\,\ref{FigFreqIncrease}(d)] but is proportional to the coupling $\kappa$ [Fig.\,\ref{FigFreqIncrease}(e)]. In addition, the receiver efficiency is maximized in the case of $\delta\omega=0$.

We have also verified numerically that the receiver efficiency is maximized when $\delta\omega=0$ for rectangular, exponentially increasing, and exponentially decreasing drive pulses, even when the pulse lengths and time constants are varied. The efficiency is also an even function of the detuning. In addition, this is even true when the intrinsic resonator loss is nonzero.

\subsection{Intrinsic Loss}

Suppose the resonator has intrinsic loss characterized by a decay time $T_1$ but the detuning is zero and the coupling is static. Then, Eq.\,(\ref{EqFullBDot}) reduces to
\[
\dot{B} = \left(-\frac{\kappa}{2} - \frac{1}{2T_1}\right)B + \frac{\mathbf{t_1}}{\tau_{rt}}\frac{\mathbf{r_2^*}}{|\mathbf{r}|}A,
\]
where the receiver efficiency can be found as before. Now suppose that the time-dependence of $A(t)$ can be expressed as a function of $t/\tau$ for some time constant $\tau$; this is valid for both rectangular and exponential pulses. Then, this efficiency incorporating $T_1$ equals that calculated from Eqs.\,(\ref{EqEnergyRes})-(\ref{EqEnergyTot}) with the following modifications:
\begin{enumerate}
\item
$T'\rightarrow\frac{\kappa+1/T_1}{\kappa}T'$
\item
$T\rightarrow\frac{\kappa+1/T_1}{\kappa}T$
\item
$\tau\rightarrow\frac{\kappa+1/T_1}{\kappa}\tau$
\item
The receiver efficiencies are reduced by $\frac{\kappa}{\kappa+\frac{1}{T_1}}$.
\end{enumerate}
For example, the receiver efficiency for an exponential pulse is
\[
\frac{E_{res}}{E_{tot}} =
\frac{4 \left(\frac{2}{\left(\kappa+\frac{1}{T_1}\right)\tau}\right) \left(e^{T/\tau}-e^{-\left(\kappa+\frac{1}{T_1}\right)T/2}\right)^2}
{\left(1+\frac{2}{\left(\kappa+\frac{1}{T_1}\right)\tau}\right)^2 \left(e^{2T'/\tau}-1\right)}
\frac{\kappa}{\kappa+\frac{1}{T_1}},
\]
which is just Eq.\,(\ref{EqExpChopped}) with these modifications. Note that $\left(\kappa+\frac{1}{T_1}\right)$ is just the measured decay constant when performing a typical $T_1$ measurement with a given $\kappa$.

\clearpage
\section{Experimental Methods}

Here we describe our experimental setup along with how we extract absorption efficiencies from the raw data. We then describe how to use the superconducting phase qubit to experimentally calibrate the couplings between the resonator and the transmission line, delay times between the various control lines, and the resonator drive energy. We demonstrate the independence of the absorption efficiency on the drive power and number of averages.

\subsection{Experimental Setup and Measurement}

\begin{figure}
\includegraphics{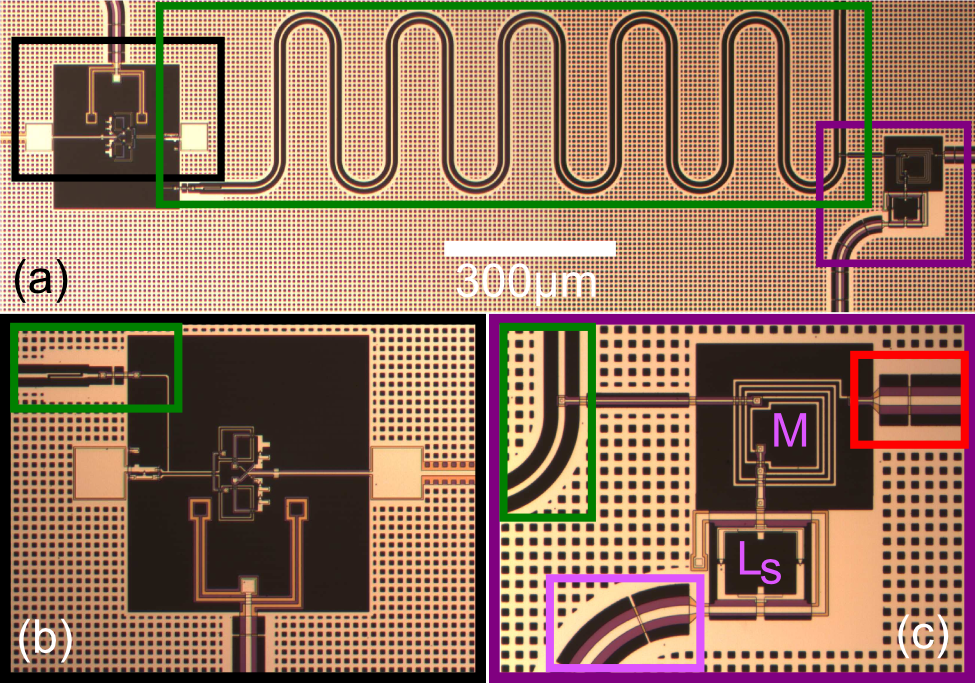}
\caption{\label{FigPhoto} Photomicrograph of experimental device, showing chip (a), superconducting phase qubit (b), and tunable coupler (c). Qubit, resonator, coupler, coupler bias line, and drive/measure line are respectively in black, green, dark purple, light purple, and red. Qubit is used for calibrating coupler.}
\end{figure}

The device is the same as was used in Ref.\,\cite{Yin2013}, as shown in Fig.\,\ref{FigPhoto}. This multi-layer device was patterned using standard photolithography \cite{Martinis2009}. All metal films were sputter-deposited Al, with \textit{in situ} Ar ion milling prior to deposition; they were etched with a BCl$_3$/Cl$_2$ inductively-coupled plasma \cite{Megrant2012}. The phase qubit and SQUID (superconducting quantum interference device) are comprised of Al/AlO$_x$/Al junctions. The phase qubit capacitor and crossovers were created using a hydrogenated amporphous silicon dielectric.

The chip with the resonator is mounted on the 30\,mK stage of a dilution refrigerator. The chip is located in an Al sample mount \cite{Wenner2011} placed in a high-permeability magnetic shield. This protects against magnetic vortex losses and prevents magnetic fields from other components, such as circulators and switches, from changing the device calibration.

\begin{figure}
\includegraphics{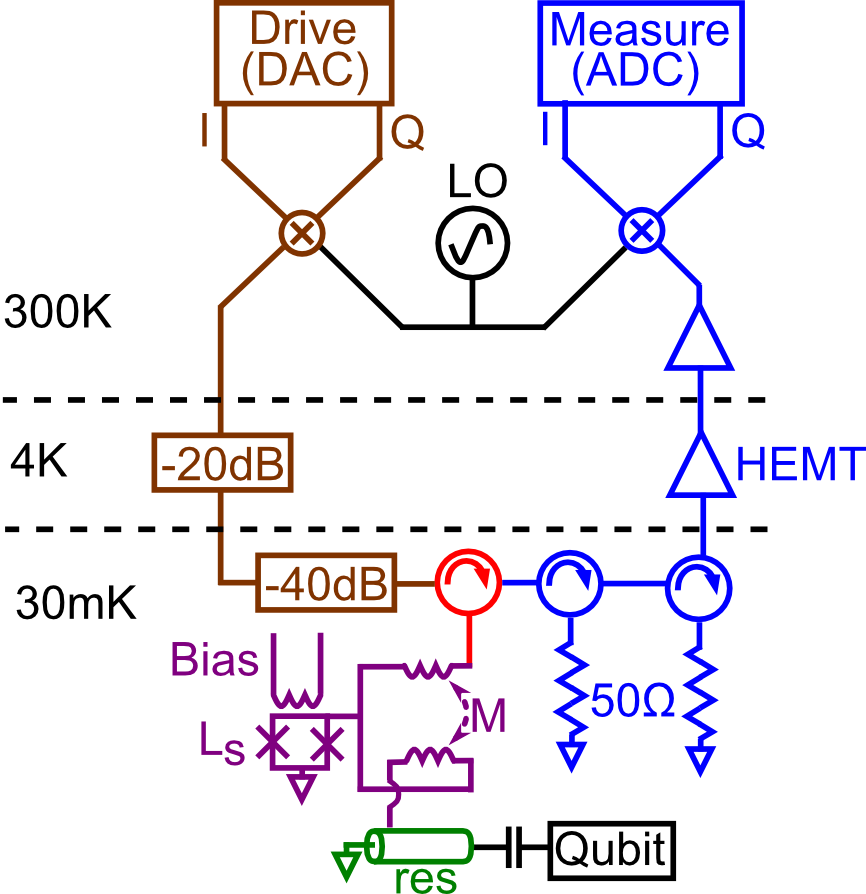}
\caption{\label{FigSchematic} Experimental schematic. The resonator is driven through a tunable coupler (consisting of a mutual inductance $M$ modulated by a SQUID with inductance $L_s$. The drive signal is generated by a digital-to-analog converter (DAC), mixed with local oscillator (LO) at 6.4\,GHz, and attenuated. Signals reflected from the coupler and leaking from the resonator are separated from the drive path using a circulator. These output signals then pass through two circulators for thermal noise isolation, are amplified at 4\,K by a HEMT amplifier and then further at room temperature, are heterodyne mixed with an $IQ$ mixer using the drive LO, and then are measured using an analog-to-digital converter (ADC).}
\end{figure}

We generate the resonator drive pulse with a digital-to-analog converter (DAC) \cite{Yin2013,Martinis2009,Chen2012,Hofheinz2009}. Each board contains two channels with a one-gigasample-per-second 14-bit DAC chip; the outputs correspond to the $I$ (cosine) and $Q$ (sine) quadratures. Both outputs are mixed by an $IQ$-mixer with a continuous $\sim$6.4\,GHz sine wave from a local oscillator (LO), as shown in the schematic of Fig.\,\ref{FigSchematic}. The LO is detuned by $f_{sb}=$165\,MHz from the resonator frequency to prevent spurious residual signal at the LO frequency from exciting the resonator; we compensate for this with the $I$ and $Q$ signals. We calibrate out mixer imperfections as explained in \cite{Hofheinz2009}.

To reach the single-photon level, we attenuate the drive signal and amplify the reflected signal, as shown in Fig.\,\ref{FigSchematic}. In particular, we attenuate themal noise with 20\,dB of attenuation at 4\,K and 40\,dB of attenuation at 30\,mK. We separate out the reflections on the drive line with a circulator. The reflections then pass through two circulators to isolate the resonator from thermal noise and are then amplified by $\sim$35\,dB at 4\,K with a low noise HEMT (high-electron mobility transistor) amplifier; there is additional amplification at room temperature.

We measure the amplified reflection signal with a room temperature analog-to-digital converter (ADC) \cite{Yin2013,Chen2012}. We first down-convert the signal with an $IQ$-mixer using the LO signal. The resulting $I$ and $Q$ quadrature voltages are then measured versus time using two 500-megasample-per-second 8-bit ADC chips and are then averaged $\sim3,000,000$ times.

We then filter the raw $V(t)=I+iQ$ data. We first rescale the $Q$ data by an experimentally measured factor to accomodate differneces in electronics between the two quadratures. We then subtract the average value of $V(t)$ as measured prior to the drive to remove DC components. We then multiply by $e^{i2\pi f_{sb}}$ to determine the signal at the drive frequency. To remove crosstalk signals, we digitally apply a sharp low-pass filter at 150\,MHz.

\begin{figure}
\includegraphics{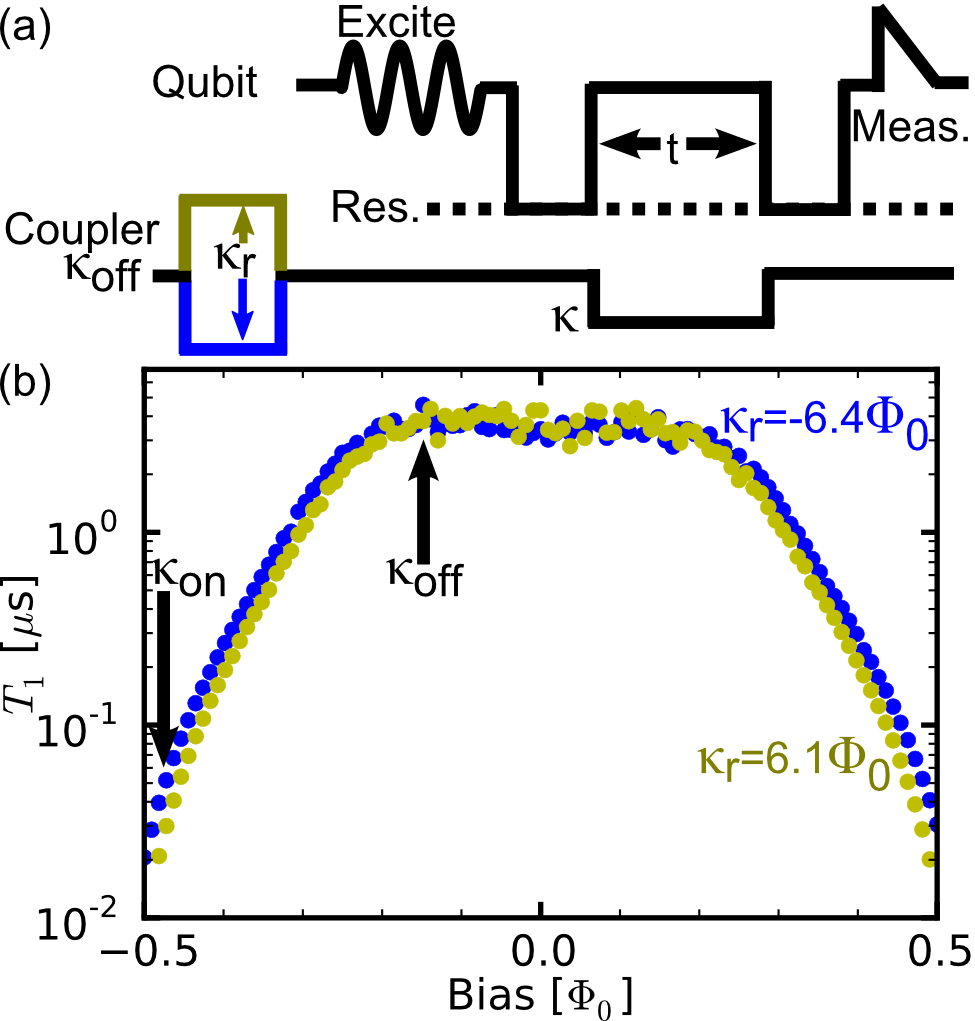}
\caption{\label{FigCoupling}
Calibrating coupling. (a) Pulse sequence. We first reset the coupler while resetting the qubit into the ground state. We then excite the qubit and swap the excitation into the resonator. After a time $t$ with the excitation in the resonator, any remaining excitation is swapped back into the qubit and then measured. (b) The decay time $T_1$ is plotted versus the coupler bias current, expressed as a flux $\Phi$ through the SQUID divided by the flux quantum $\Phi_0=h/2e$. Decay times are shown for two coupler reset biases, $+6.1\,\Phi_0$ (yellow) and $-6.4\,\Phi_0$ (blue). With the $-6.1\,\Phi_0$ bias, we define the coupling $\kappa_{\mbox{on}}$ to be on when $T_1=1/50$\,ns; this is where we drive the resonator and allow retransmission. When $T_1$ is maximized, the coupling $\kappa_{\mbox{off}}$ is  zero and so is turned off.}
\end{figure}

In calculating the energy $\int |V(t)|^2\,dt$ in each portion of the pulse sequence, any noise appears to be additional energy. To subtract out this spurious contribution, our procedure (as rigorously derived in the ``Error Analysis'' section) involves:
\begin{itemize}
\item Calculate the energy prior to driving the resonator, where there is no signal
\item Rescale the noise energy to determine the noise energy in the desired integration region
\item Subtract this noise energy from the total measured energy in a desired region.
\end{itemize}

\subsection{Calibrating Coupling \& Delays}

The tunable coupler employs a tunable mutual inductance between the drive and resonator \cite{Yin2013}. The mutual inductance $M$ consists of two interwound coils which are galvanically connected, as shown in Fig.\,\ref{FigPhoto}(c) and Fig.\,\ref{FigSchematic}. From this connection, a SQUID (superconducting quantum interference device) is attached with tuning inductance $L_s$. This tuning arises from a flux induced by an on-chip coupler bias current which is externally generated.

\begin{figure}
\includegraphics{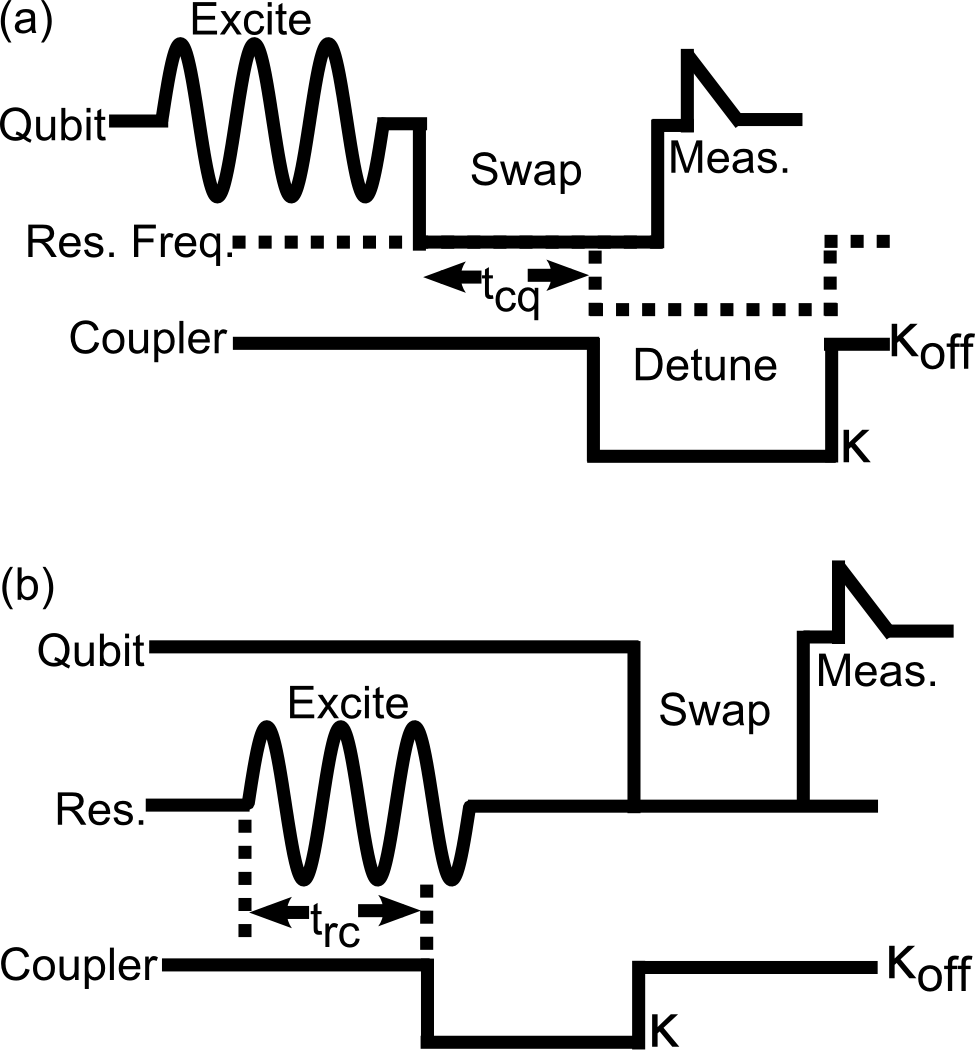}
\caption{\label{FigTestDelays}
Correcting for inter-line delays. (a) To measure the coupler-qubit delay $t_{cq}$, we excite the qubit, swap the excitation into the resonator detuned by opening the coupler, and measure the qubit. (b) To measure the resonator-coupler delay $t_{rc}$, we drive the resonator and open the coupler for the same duration but offset by $t_{rc}$. We then swap any resonator excitation to the qubit, which is then measured.}
\end{figure}

\begin{figure*}
\includegraphics{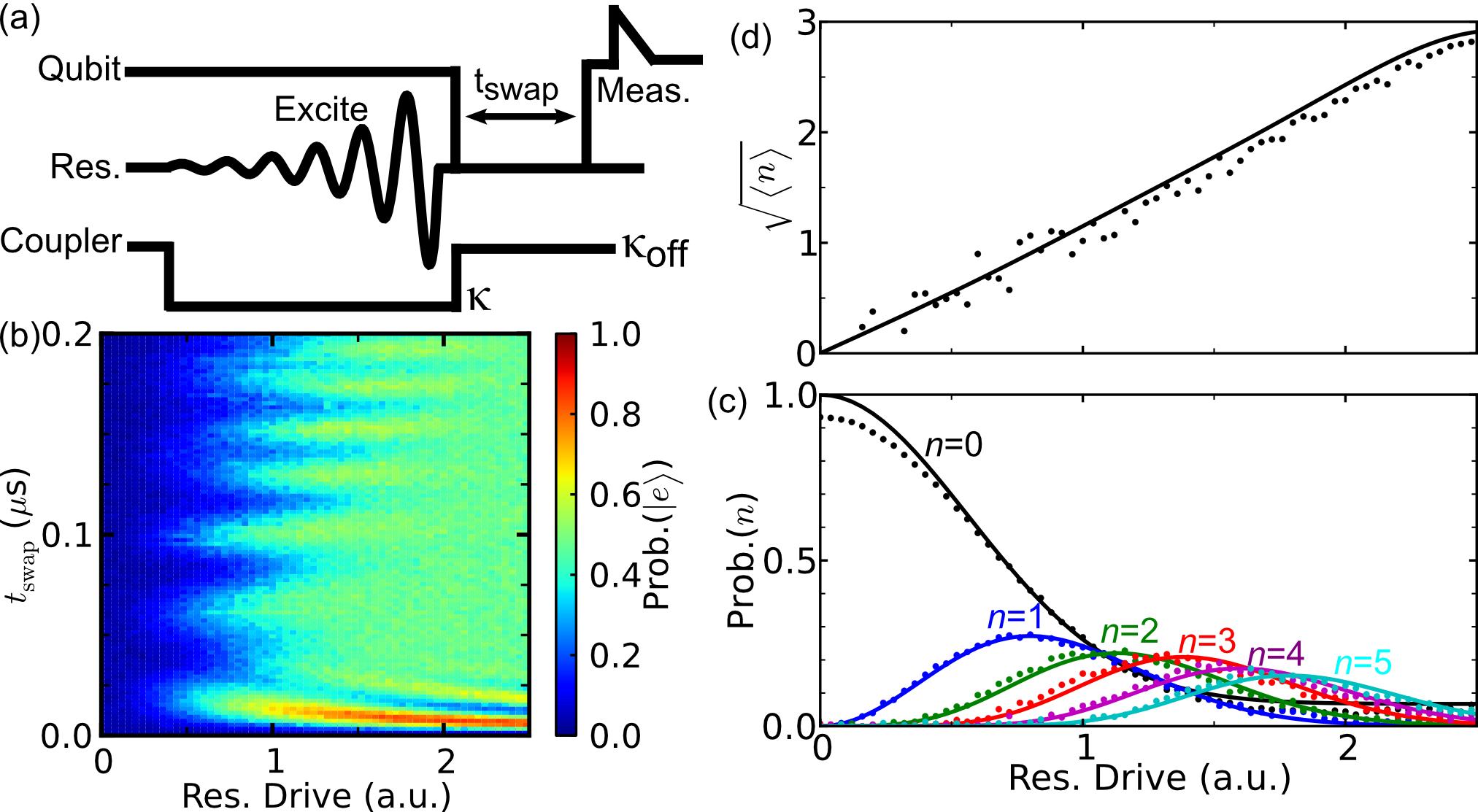}
\caption{\label{FigCoherent}
Calibrating drive energy. (a) With the coupler open, we drive the resonator with various drive amplitudes. We then swap any excitation to the qubit for varying times $t_{\textrm{swap}}$ and measure the qubit. (b) The qubit excited state ($|e\rangle$) probability is shown versus the drive amplitude and $t_{\textrm{swap}}$. For these pulses, we used a 1\,$\mu$s exponentially increasing pulse with $\tau=2/\kappa$. (c) Probability for the resonator to be in the $n$th Fock state, fit to the data in (b), versus drive amplitude. States with $n>5$ are not shown. (d) The average photon number from the probabilities in (c) is shown versus resonator drive amplitude.}
\end{figure*}

We calibrate the coupling $\kappa\propto(M-L_s)^2$ \cite{Yin2013} in terms of this current with a superconducting phase qubit capacitively coupled to the resonator \cite{Yin2013,Neeley2008,Wang2008}. Using the qubit, we generate a single photon and swap it to the resonator [Fig.\,\ref{FigCoupling}(a)]. We then apply the desired coupler bias for varying times, after which we swap to the qubit and measure the residual excitation. From the decay of the excitation probability $P_e$, we extract the resonator decay time $T_1$, shown in Fig.\,\ref{FigCoupling}(b). At the bias maximizing $T_1$, the coupling is zero and hence closed, as verified by trying to drive the resonator; here, $T_1$ is the intrinsic resonator decay time $1/\kappa_i\simeq3\,\mu$s since $1/T_1=\kappa+\kappa_i$. We define the coupler to be open and drive the resonator when $T_1\simeq50$\,ns.

However, the open coupling can range from 1/(50\,ns) to 1/(30\,ns), since the SQUID has multiple potential wells, each with a different coupling (see Fig.\,\ref{FigCoupling}(b)). We reproducibly select a particular well by adding a coupler reset pulse prior to all pulse sequences.

Adjusting the SQUID to tune the coupler modulates the resonator inductance to ground \cite{Yin2013}. This adjusts the resonator frequency by $\sim20$\,MHz between the opened and closed biases. Hence, opening the coupler to detune the resonator blocks swaps between the qubit and resonator as tuned with the coupler closed.

We use this to tune the temporal delay $t_{cq}$ between the qubit and coupler [Fig.\,\ref{FigTestDelays}(a)]. We first excite the qubit, swap the excitation to the resonator while opening the coupler, and measure the qubit. As we vary $t_{cq}$, the time for which the qubit excited state probability $P_e$ is maximized is the actual delay.

To calibrate the delay $t_{rc}$ between the resonator drive and the coupler, we employ the sequence in Fig.\,\ref{FigTestDelays}(b). Here, we drive the resonator with a many-photon pulse to ensure resonator excitation. For this time but offset by $t_{rc}$, we open the coupler. Any induced resonator excitation is then swapped into the qubit and measured. Hence, the actual delay is when $P_e$ is maximized. As explained in the main text, we verify this timing by varying when the coupler is closed relative to stopping the drive and maximizing the absorption efficiency.

\subsection{Drive Energy}

We use the qubit to calibrate the resonator drive energy in terms of the drive amplitude \cite{Yin2013, Hofheinz2008}. We vary the drive amplitude for a particular drive pulse shape. For each amplitude, we drive the resonator, swap between the qubit and resonator for varying swap times $t_{\mbox{swap}}$, and measure the qubit excited state probability $P_e$ [Fig.\,\ref{FigCoherent}(a),(b)]. We simulate this probability versus $t_{\mbox{swap}}$ using the Linblad master equation for $n$-photon Fock states. We least-squares fit the experimental and theoretical probabilities to determine for each drive amplitude the experimental Fock state distribution, shown in Fig.\,\ref{FigCoherent}(c). We fit this measured distribution to a Poisson distribution,
\[
P_n^\textrm{Poisson} = \frac{\langle n\rangle^n e^{-\langle n\rangle}} {n!},
\]
as the resonator is in a classically-driven coherent state. From this fit, we extract the mean number of photons $\langle n\rangle$ captured in the resonator [Fig.\,\ref{FigCoherent}(d)] and find a linear fit between drive amplitude and $\sqrt{\langle n \rangle}$. We then rescale this according to the measured absorption efficiency to determine the drive amplitude necessary for a single photon drive ($\langle n \rangle=1$).

\begin{figure}
\includegraphics{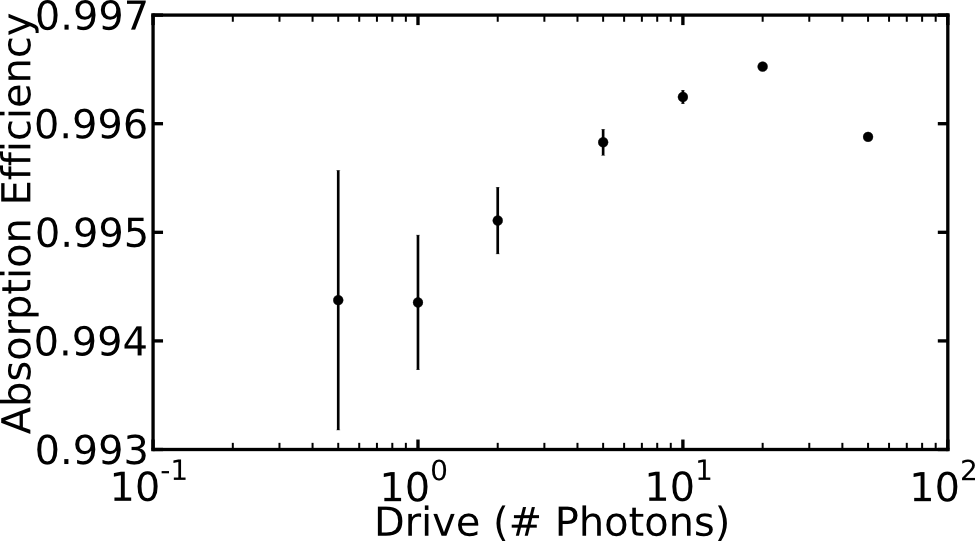}
\caption{\label{FigDrivePower}
Drive energy independence of absorption efficiency. The absoprtion efficiency is independent of the resonator drive energy for energies of 0.5-50 photons. Data are shown for exponentially increasing drive pulses with time constant $\tau=2/\kappa$ and length $20/\kappa$, and error bars indicate statistical errors from Eq.\,(\ref{EqUncertaintyAbs}).}
\end{figure}

To determine if the exact calibration matters, we measured the absorption efficiency versus resonator drive amplitude. As shown in Fig.\,\ref{FigDrivePower}, the capture efficiency is independent of the drive energy between 0.5 and 50 photons. This demonstrates that, although the theoretical capture efficiencies were calculated in the classical limit, they are still valid in the quantum regime. We note that even if the energy calibration is mistuned, the absorption efficiencies quoted in the main text are still valid.

\subsection{Averaging}

\begin{figure}
\includegraphics{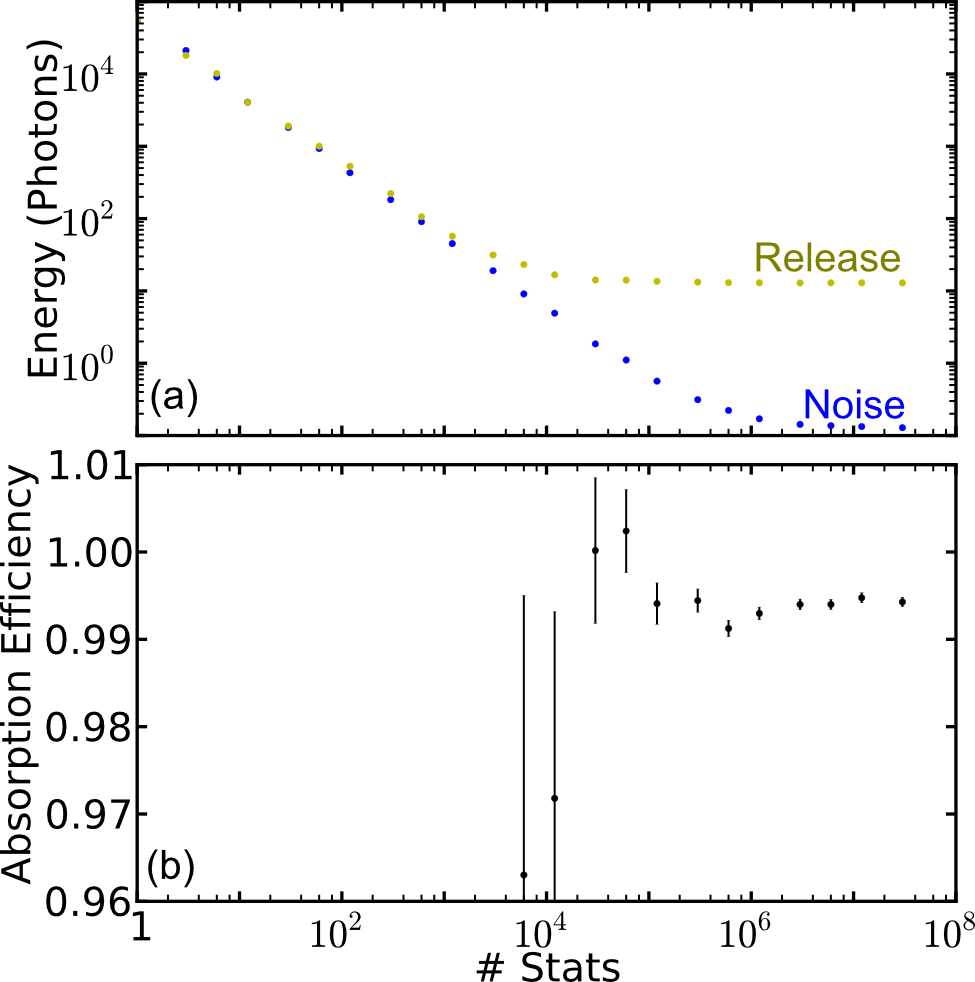}
\caption{\label{FigNumAverages}
Varying number of averages. (a) The noise (blue) and release signal (yellow) energies are shown for varying numbers of averages. The noise energy has been rescaled by the duration of the energy release phase of the pulse sequence (See Fig.\,2 of the main text). These are measured for exponentially increasing drive pulses with time constant $\tau=2/\kappa$ and length $20/\kappa$. The release energy flattens out where it no longer equals the noise energy. (b) Absorption efficiencies versus number of averages. Data are only shown where the noise energy is substantially less than the release energy so that absorption efficiencies make sense. Error bars indicate statistical errors from Eq.\,(\ref{EqUncertaintyAbs}) and are approximately 6\% the noise-to-signal ratio.}
\end{figure}

We also checked whether the number of averages affects the absorption efficiencies. As the number of averages is increased, the signal amplitude would tend to zero if the signal lacked phase coherence. However, as shown in Fig.\,\ref{FigNumAverages}(a), the signal energy instead is constant once the signal dominates over the noise. Similarly, the absorption efficiencies [Fig.\,\ref{FigNumAverages}(b)] are nearly constant in this regime but are frequently unphysical when the noise is dominant.

The noise energy scales inversely with the number of averages. This makes sense as the energy is proportional to the square of the measured voltage. For large numbers of averages, the noise energy becomes constant; we measure the absorption efficiencies near the start of this regime to ensure maximal signal-to-noise ratio. The uncertainties in the absorption efficiency, as calculated with Eq.\,(\ref{EqUncertaintyAbs}), are approximately 6\% the ratio of the release phase noise energy to the total signal energy.

\clearpage
\section{Error Analysis}

Here we explain how to subtract noise to get unbiased estimates of energies. We then calculate the random errors in terms of experimental quantities and consider possible sources of systematic errors.

\subsection{Noise Subtraction}

We calculate the energies for the absorption efficiency by integrating the magnitude-squared of the signal; however, noise contributes to these calculated energies. Assume that, at the $k$th time step, the actual signal is $(I_k,Q_k)$, and the noise is $(x_k,y_k)$ in the $I$- and $Q$-quadratures, respectively. Then, if the duration of each time step is $t$ and there are $N$ time steps, the calculated energy is
\[
E_{sig} = t \sum_{k=1}^N [(I_k+x_k)^2 + (Q_k+y_k)^2].
\]
The energy in the absence of noise is
\begin{equation}
t \sum_{k=1}^N [I_k^2 + Q_k^2].
\label{EqNoiseless}
\end{equation}
With no signal, the energy is the noise energy 
\[
E_{N} = t \sum_{k=1}^N [x_k^2 + y_k^2],
\]
so noise energy is measured even without reflections. Hence, the directly measured capture efficiencies are lower than the true values.

To remove this noise contribution, we assume the noise in each quadrature is Gaussian and uncorrelated with noise in the other quadrature, noise at other times, and the signal amplitude. These assumptions are consistent with wide bandwidth white noise in our experiment. We further assume the noise has standard deviation $\sigma$ and zero mean. With these assumptions, the noise for time step $k$ can be treated as random values $g_k$ which are independent with identical Gaussian distributions. Averaging over all such random values gives moments
\[
\langle g_k^p\rangle = \frac{1+(-1)^p}{2} (p-1)!!\sigma^p.
\]
Particularly useful moments are $\langle g_k\rangle=\langle g_k^3\rangle=0$, $\langle g_k^2\rangle=\sigma^2$, and $\langle g_k^4\rangle=3\sigma^4$.

Before considering the statistical properties of these energies, we first consider the first and second moments of two useful sums. The first is a weighted sum of Gaussian distributed noise signals given by $\sum_{k=1}^N w_kg_k$, where $w_k$ is the weight of the $k$th time step. By the linearity of expectation values,
\[
\left\langle\sum_{k=1}^N w_kg_k\right\rangle = 0,
\]
while the second moment is
\begin{eqnarray*}
\left\langle\left(\sum_{k=1}^N w_kg_k\right)^2\right\rangle
&=& \sum_{k=1}^N \sum_{l=1}^N w_kw_l\langle g_kg_l\rangle \nonumber\\
&=& \sum_{k=1}^N w_k^2\langle g_k^2\rangle + \sum_{k\neq l} w_kw_l\langle g_k\rangle\langle g_l\rangle \nonumber\\
&=& \sigma^2 \sum_{k=1}^N w_k^2,
\end{eqnarray*}
where we use the Gaussian moments and the independence of noise signals at different times. The other useful sum is a weighted sum of the square of Gaussian distributed noise signals, $\sum_{k=1}^N w_kg_k^2$ for weights $w_k$. By the linearity of expectation values,
\[
\left\langle\sum_{k=1}^N w_kg_k^2\right\rangle = \sigma^2\sum_{k=1}^N w_k.
\]
The second moment of this sum is given by
\begin{eqnarray*}
\left\langle\left(\sum_{k=1}^N w_kg_k^2\right)^2\right\rangle
&=& \sum_{k=1}^N \sum_{l=1}^N w_kw_l\langle g_k^2g_l^2\rangle \nonumber\\
&=& \sum_{k=1}^N w_k^2\langle g_k^4\rangle + \sum_{k\neq l} w_kw_l\langle g_k^2\rangle\langle g_l^2\rangle \nonumber\\
&=& 3\sigma^4 \sum_{k=1}^N w_k^2 + \sigma^4 \sum_{k\neq l} w_kw_l \nonumber\\
&=& 2\sigma^4 \sum_{k=1}^N w_k^2 + \sigma^4 \left(\sum_{k=1}^N w_k\right)^2.
\end{eqnarray*}

We can now calculate the mean of $E_{sig}$ and derive how to subtract the noise contribution. The means of $E_{sig}$ and $E_N$ equal
\begin{eqnarray}\label{EqESMean}
\langle E_{sig}\rangle &=& t\left[ \sum_{k=1}^{N_S} (I_k^2+Q_k^2) + N_S\sigma_x^2 + N_S\sigma_y^2 \right]\\
\langle E_N\rangle &=& t\left[ N_N\sigma_x^2 + N_N\sigma_y^2 \right]
\label{EqENMean}
\end{eqnarray}
where the signal (noise) is measured with $N_S$ ($N_N$) time steps and the noise in the $I$ ($Q$) quadrature has standard deviation $\sigma_x$ ($\sigma_y$). With $E_N$ rescaled by $N_S/N_N$, these means differ by the noiseless energy, so we remove the contribution of noise in a region by:
\begin{itemize}
\item Calculate $E_N$ prior to driving the resonator, so $I_k=Q_k=0$.
\item Rescale $E_N$ by the duration of the desired region.
\item Subtract this noise energy from the total measured energy.
\end{itemize}
This gives an energy
\[
E_{sig}^{NS} = E_{sig} - \frac{N_S}{N_N}E_N,
\]
which is an unbiased estimate of the noiseless energy [Eq.\,(\ref{EqNoiseless})].

\subsection{Error Analysis}

To determine the random uncertainty from this noise, we calculate the variance of both $E_{sig}$ and $E_{sig}^{NS}$. We assume that the following pairs are uncorrelated: noise at different times, noise in the two quadratures, and noise with the signal amplitude. Using the Gaussian moments calculated earlier, we find the following variances:
\begin{eqnarray*}
\langle(\Delta E_{sig})^2\rangle &=& 4t^2\sum_{k=1}^{N_S}(I_k^2\sigma_x^2 + Q_k^2\sigma_y^2) + 2N_St^2(\sigma_x^4 +\sigma_y^4) \\
\langle(\Delta E_N)^2\rangle &=& 2N_Nt^2(\sigma_x^4 + \sigma_y^4).
\end{eqnarray*}
Since $E_{sig}$ and $E_N$ are measured at different times and noise at different times is uncorrelated, $E_{sig}$ and $E_N$ are uncorrelated and so the variance in $E_{NS}$ equals
\[
\langle(\Delta E_{sig}^{NS})^2\rangle = \langle(\Delta E_{sig})^2\rangle + \left(\frac{N_S}{N_N}\right)^2\langle(\Delta E_N)^2\rangle.
\]
The additional uncertainty scales as $\sigma^4$ while the uncertainty in the raw signal energy scales as $\sigma^2$. With a large signal-to-noise ratio, the additional noise from this procedure can be neglected.

Further, these variances can be reexpressed in terms of the measured energies $\langle E_{sig}^{NS}\rangle$ and $\langle E_N\rangle$ using Eqs.\,(\ref{EqESMean})-(\ref{EqENMean}). In particular, if $\sigma_x=\sigma_y$,
\begin{eqnarray}
\langle(\Delta E_{sig})^2\rangle &=& \frac{2}{N_N}\langle E_{sig}^{NS}\rangle\langle E_{N}\rangle + \frac{N_S}{N_N^2}\langle E_{N}\rangle^2 \nonumber\\
\langle(\Delta E_{sig}^{NS})^2\rangle &=& \frac{2}{N_N}\langle E_{sig}^{NS}\rangle\langle E_{N}\rangle + \frac{N_S(N_S+N_N)}{N_N^3}\langle E_{N}\rangle^2 \nonumber\\
\langle(\Delta E_{N})^2\rangle &=& \frac{1}{N_N}\langle E_{N}\rangle^2.
\label{EqENSVariance}
\end{eqnarray}

We then use these expressions to determine the uncertainty in the absorption efficiency, the ratio of the energy $E_{abs}^{NS}$ absorbed and then released by the resonator to the total measured energy $E_{tot}^{NS}$. However, $E_{tot}^{NS}=E_{abs}^{NS}+E_{ref}^{NS}$ for reflected energy $E_{ref}^{NS}$, so the uncertainties in $E_{abs}^{NS}$ and $E_{tot}^{NS}$ are not independent. Similarly, all noise-subtracted energies contain the single term $\langle E_{N}\rangle$ and so have correlated uncertainties. Since $E_{abs}$, $E_{ref}$, and $E_N$ are measured at different times, these energies are independent, so their uncertainties can be used to calculate the overall absorption efficiency uncertainty
\begin{widetext}
\begin{equation}
\delta\left(\frac{E_{abs}^{NS}}{E_{tot}^{NS}}\right) =
\sqrt{\langle(\Delta E_{abs})^2\rangle\left(\frac{E_{ref}^{NS}}{\left(E_{tot}^{NS}\right)^2}\right)^2
+\frac{\langle(\Delta E_{ref})^2\rangle}{\left(E_{tot}^{NS}\right)^4}
+\langle(\Delta E_{N})^2\rangle\left(\frac{N_{ref}E_{abs}^{NS}-N_{abs}E_{ref}^{NS}}{N_N\left(E_{tot}^{NS}\right)^2}\right)^2},
\label{EqUncertaintyAbs}
\end{equation}
\end{widetext}
where $N_{ref}$ and $N_{abs}$ are the number of data points used to measure $E_{ref}$ and $E_{abs}$, respectively. To verify this uncertainty, we repeated the measurement of the minimal-reflection absorption efficiency 60 times and measured a standard deviation in the absorption efficiency of 0.0552\%, within 1\% of the 0.0548\% expected according to Eq.\,(\ref{EqUncertaintyAbs}), validating this error analysis.

The storage efficiency is the ratio of $E_{abs}^{NS}$ to the total pulse energy $E_{off}^{NS}$ measured with the coupler off. Since these signals and the noise contributions are measured in different experiments, the uncertainties in these energies are independent, so standard error propagation applies.

These uncertainties only cover random variations but not systematic errors. One major source of systematic errors is poor signal or drive path calibration; this is a multiplicative effect and so changes the raw energies but not ratios such as the absorption and receiver efficiencies. These efficiencies can, however, be reduced by imperfections in the pulse calibration. We scan over coupler closing delay and drive frequency, measure the resulting absorption efficiencies for an exponentially increasing pulse, and choose parameters which maximize the absorption efficiency. However, this does not include imperfections in the pulse shape, which are likely reducing our measured efficiency as explained in the discussion of the coupler delay in the main text.